# Searching for C II Emission from the First Sample of $z \sim 6$ O I Absorption-associated Galaxies with the Atacama Large Millimeter/submillimeter Array

Yunjing Wu[1,2], Zheng Cai[1], Jianan Li[1], Kristian Finlator[3,4], Marcel Neeleman[5], J. Xavier Prochaska[6,7],
Bjorn H. C. Emonts[8], Shiwu Zhang[1], Feige Wang[2], Jinyi Yang[2,13], Ran Wang[9], Xiaohui Fan[2], Dandan Xu[1],
Emmet Golden-Marx[1], Laura C. Keating[10], and Joseph F. Hennawi[11,12]
[1] Department of Astronomy, Tsinghua University, Beijing 100084, People's Republic of China; zcai@mail.tsinghua.edu.cn
[2] Steward Observatory, University of Arizona, 933 N Cherry Avenue, Tucson, AZ 85721, USA
[3] New Mexico State University, Las Cruces, 88003 NM, USA
[4] Cosmic Dawn Center (DAWN), Niels Bohr Institute, University of Copenhagen/DTU-Space, Technical University of Denmark, Denmark
[5] Max-Planck-Institut für Astronomie, Königstuhl 17, D-69117, Heidelberg, Germany
[6] Department of Astronomy and Astrophysics, UCO/Lick Observatory, University of California, 1156 High Street, Santa Cruz, CA 95064, USA
[7] Kavli Institute for the Physics and Mathematics of the Universe (Kavli IPMU), The University of Tokyo, 5-1-5 Kashiwanoha, Kashiwa, 277-8583, Japan
[8] National Radio Astronomy Observatory, 520 Edgemont Road, Charlottesville, VA 22903, USA
[9] Department of Astronomy, Peking University, Beijing 100871, People's Republic of China
[10] Institute for Astronomy, University of Edinburgh, Blackford Hill, Edinburgh, EH9 3HJ, UK
[11] Department of Physics, University of California, Santa Barbara, CA 93106-9530, USA
[12] Leiden Observatory, Leiden University, PO Box 9513, NL-2300 RA Leiden, The Netherlands
Received 2022 July 2; revised 2023 September 18; accepted 2023 October 4; published 2023 November 8

## Abstract

We report the first statistical analyses of [C II] and dust continuum observations in six strong O I absorber fields at the end of the reionization epoch obtained by the Atacama Large Millimeter/submillimeter Array (ALMA). Combined with one [C II] emitter reported in Wu et al., we detect one O I-associated [C II] emitter in six fields. At redshifts of O I absorbers in nondetection fields, no emitters are brighter than our detection limit within impact parameters of 50 kpc and velocity offsets between $\pm 200$ km s$^{-1}$. The averaged [C II]-detection upper limit is $<0.06$ Jy km s$^{-1}$ (3$\sigma$), corresponding to the [C II] luminosity of $L_{[C\,II]} < 5.8 \times 10^7 L_\odot$ and the [C II]-based star formation rate of SFR$_{[C\,II]} < 5.5\ M_\odot$ yr$^{-1}$. Cosmological simulations suggest that only $\sim 10^{-2.5}$ [C II] emitters around O I absorbers have comparable SFR to our detection limit. Although the detection in one out of six fields is reported, an order of magnitude number excess of emitters obtained from our ALMA observations supports that the contribution of massive galaxies that caused the metal enrichment cannot be ignored. Further, we also found 14 tentative galaxy candidates with a signal-to-noise ratio of $\approx 4.3$ at large impact parameters ($>50$ kpc) and having larger outflow velocities within $\pm 600$ km s$^{-1}$. If these detections are confirmed in the future, then the mechanism of pushing metals at larger distances with higher velocities needs to be further explored from the theoretical side.

*Unified Astronomy Thesaurus concepts:* High-redshift galaxies (734); Quasars (1319)

## 1. Introduction

Cosmological reionization occurs when hydrogen transitions from its neutral to ionized state in the early universe. Investigating the behavior of neutral hydrogen (H I) can help to reveal the astrophysics driving reionization and its timing. Unfortunately, H I in the circumgalactic medium (CGM) is quite difficult to observe because of the Gunn–Peterson effect (Gunn & Peterson 1965), the nearly complete absorption of photons with rest-frame excitation energy higher than that of Ly$\alpha$ by the intergalactic medium (IGM). Moreover, the absorption caused by the gas in the CGM, with $\lambda_{\rm rest} < 1216$ Å, will also fall into the Gunn–Peterson trough and will blanket in the Ly$\alpha$-caused absorption and then be undetectable (Simcoe et al. 2020). Thus, alternative tracers are required. Because neutral oxygen has a similar ionization energy as H I, it is regarded as a possible tracer for cosmological reionization (Oh 2002; Finlator et al. 2013; Doughty & Finlator 2019).

Metal absorption is commonly observed in QSO spectra. Thanks to the rapid development of the recent high$-z$ QSO surveys (e.g., Bañados et al. 2016; Wang et al. 2019; Yang et al. 2019), $\sim 260$ QSOs have been identified at $z = 6$–7.5, which enables us to search for metal absorbers systematically (e.g., Becker et al. 2019; Cooper et al. 2019; Zou et al. 2021). Additionally, the early universe is proposed to be in a metal-poor state. Therefore, at high redshift ($z \gtrsim 6$), the existence of metal absorption systems (e.g., O I, Mg II, and C IV) constrains the nature and location of the source galaxies that contribute to the early-metal enrichment. Then, connecting galaxies with the gaseous reservoirs plays a crucial role in understanding galaxy formation at the end of the epoch of reionization.

Cosmological simulations suggest that galactic winds from typical star-forming galaxies can eject metals into the CGM/IGM (e.g., Keating et al. 2014; Pallottini et al. 2014; Davé et al. 2016). Inspired by these works, star formation rates (SFRs) and impact parameters of source galaxies are two key parameters that need to be measured to test models of galaxy formation. As such, the direct imaging of absorption-associated galaxies is necessary because it provides measurements of these two parameters simultaneously. Díaz et al. (2011) found a C IV

---

[13] Strittmatter Fellow.

**Table 1**
O I Absorber and QSO Information

| QSO Field Name (1) | $z_{OI}$ (2) | $\log(N_{OI}/cm^{-2})$ (3) | $t_{on}$ (hr) (4) |
|---|---|---|---|
| J2054−0005 | 5.978 | 14.2 | 2.3 |
| J2315−0023 | 5.7529 | 14.5 | 2.8 |
| J0100+2802 (F1) | 6.144 | 14.7 | 2.5 |
| J0100+2802 (F2) | 6.112 | 14.4 | 2.5 |
| PSO J183+05 | 6.064 | 14.4 | 2.0 |
| PSO J159-02 | 6.238 | 14.5 | 1.4 |

**Notes.** Columns: (1) QSO field name: F1 is the first absorber field, while F2 is the second one from the same sightline; (2) redshift of O I absorber; (3) O I column density; (4) on-source time.

absorption-associated Lyman $\alpha$ emitter (LAE) at $z = 5.791$ with Ly$\alpha$-based SFR of 1.4 $M_\odot$ yr$^{-1}$ and at the distance of 79 kpc. Additionally, Díaz et al. (2014, 2015) observed LAEs around a sample of C IV absorbers with imaging and follow-up spectroscopic analysis with the detection limit of SFR$_{Ly\alpha} \approx 5\,M_\odot$ yr$^{-1}$. They surmised that the LAE distribution may potentially trace large-scale outflows. To further constrain the possibility of faint sources as C IV absorber host galaxies, Cai et al. (2017) used Hubble Space Telescope (HST) narrow band-imaging observations to constrain the detection limit down to 2 $M_\odot$ yr$^{-1}$. Moreover, Díaz et al. (2021) used the Very Large Telescope and Multi Unit Spectroscopic Explorer (MUSE) to search for LAEs around 11 C IV absorbers and found these LAEs have impact parameters ranging from 11–200 kpc and Ly$\alpha$ luminosity of 0.18–1.15L$^*_{Ly\alpha}$. However, Ly$\alpha$ can easily be scattered due to the resonant scattering effect (e.g., Dijkstra et al. 2006; Zheng et al. 2010), which causes the Ly$\alpha$-derived star formation rates to be uncertain.

Due to the limitation of treating Ly$\alpha$ emission as galaxy tracers, only a few candidate galaxies in the field of absorbers have been identified. Because [C II]-158 $\mu$m emission is one of the best ISM indicators (e.g., Wang et al. 2013; Decarli et al. 2017; Neeleman et al. 2019), it is an alternative tracer at $z \sim 6$. To trace H I at $z \gtrsim 6$, we choose to use O I absorption as an alternative. As such, we develop our approach following this methodology. To further investigate the connection between absorbers and their host galaxies, we use the Atacama Large Millimeter/submillimeter Array (ALMA) to search for O I absorber-[C II] emitter pairs in QSO fields.

We chose to use ALMA because it also provides us with deep continuum observations, which will allow us to investigate the source densities/environment of these high-$z$ QSOs. Previously, Wu et al. (2021) reported the results in the field of QSO J2054−0005. Here, we follow up on work and present observations of the additional four fields. These five systems also form the first sample to study metal absorber–submillimeter galaxy interactions. Simulations predict that QSOs at $z \sim 6$ with a black hole mass of $M_{BH} \approx 10^9 M_\odot$ are primarily embedded in massive halos and located in overdense regions (Costa et al. 2014). Observationally, Decarli et al. (2017) and Trakhtenbrot et al. (2017) conducted ALMA observations in QSO fields and found several continuum sources and [C II] companions at the redshifts of QSOs. After comparing blind-field number counts, Decarli et al. (2017) concluded that the cumulative number of companion galaxies are excesses in QSO fields. Neeleman et al. (2019) observed more QSOs at this redshift and then obtained the same conclusion. Furthermore, at $z \sim 4$, García-Vergara et al. (2022) detected five CO(4–3) emitters around 17 QSOs, while only 0.28 CO(4–3) detections are expected with the same volume, suggesting that QSO fields have clustering properties with emitters. Conversely, except for line emitters, Champagne et al. (2018) used ALMA to search for continuum sources in 35 bright QSOs fields at $6 < z < 7$ and found no spatial over-abundances at the scale of (<1 cMpc). Conducting our ALMA observations, we can further discuss the continuum over-abundances in QSO fields.

This paper is organized as follows. In Section 2, we describe the composition of our sample, data reduction, and source-detection details. In Section 3, we present the [C II]-intensity, continuum map and compare our observations with simulations. In Section 4, we provide further discussions about the physics behind our observations. In this paper, we assume a flat cosmological model with $\Omega_M = 0.3$, $\Omega_\lambda = 0.7$, and H$_0$ = 70 km s$^{-1}$ Mpc$^{-1}$, 1″ = 5.7 kpc at $z = 6$.

## 2. Observations, Data Reduction, and Source Detection

### 2.1. Survey Description

We used ALMA in its compact configuration (C43-1) to search for [C II] 158 $\mu$m emission of absorber-associated galaxies at $z \sim 6$ (ALMA 2017.1.01088.S, 2019.1.00466.S; PI: Cai). Our sample is composed of five QSO fields containing six strong O I absorbers with log$(N_{OI}/cm^{-2}) > 14$ (Becker et al. 2011; Cooper et al. 2019). The detailed redshift, column density, and source time are collected in Table 1. Each individual ALMA observation was done using four 1.875 GHz spectral windows (SPWs). In general, in each spectral tuning, we use one SPW to center on the [C II] emission at the redshift of the O I absorber, while the remaining SPWs were used to obtain a continuum image of the field. Specifically, in the field QSO J0100+2802, because two absorbers are located at very similar redshifts, we used our SPW-setting strategy to observe [C II] emission. Our sample contains five (six) QSO (absorber) fields. Combined with Wu et al. (2021), we describe the observations of all five QSO fields in this paper.

### 2.2. Data Reduction

We reduced our data following the standard steps, which are part of the Common Astronomy Software Application (CASA v.5.6.1-8; McMullin et al. 2007), and calibrated the data based on the archival calibration script supplied by ALMA. The absolute flux uncertainties are expected to be less than 10%. After the calibration, we generated continuum images using tclean, where we excluded frequencies at the redshift of expected [C II] emission and used a natural weighting (Li et al. 2021; Wu et al. 2021). Further, to obtain emission-line data cubes, we subtracted the continuum from the data using the task uvcontsub on the line-free channels with zero-order functions. Next, all [C II]-intensity images are cleaned down to the 5$\sigma$ level. For reaching a signal-to-noise ratio (S/N) of S/N > 5 of the [C II] emission line over an entire line width ($\approx$200 km s$^{-1}$), we need to reach an S/N > 3 over 1/3 of the source line width ($\approx$66 km s$^{-1}$). Thus, our reduction procedure yields a channel width of $\approx$66 km s$^{-1}$.

### 2.3. Source-detection Algorithm

We use the source-detection Python package `DAOStar-Finder` (Bradley et al. 2020), to search for point sources on an input image with a given shape similar to a defined 2D Gaussian kernel. The adopted shape is defined by the synthesized beams in each observation. We set a threshold for a tentative detection at the $4\sigma$ level, which corresponds to the peak S/N. The standard deviation of each image is defined by the pixel-to-pixel fluctuation and calculated using the Python Code `Qubefit` (Neeleman et al. 2020). We note that, in each field, no extended sources defined as spatially larger than a synthesized beam are found in our data cube. Thus, tentative sources are all point-like, and our detection algorithm is appropriate. The fidelity is further estimated in Appendix A.

## 3. Results

### 3.1. O I Asorber-associated [C II] Emitters Sample

#### 3.1.1. [C II] Emitter Detection

Our primary science goal is to survey the [C II] emission from galaxies that could host strong O I absorbers. We strictly followed a popular method called `FINDCLUMPS` (Walter et al. 2016; Decarli et al. 2020; González-López et al. 2020). The basic idea of this algorithm is to detect 2D sources on different moment-zero maps with different line widths at different frequencies. For the first step, the potential [C II]-intensity images are generated by floating averages of a given number of channels with different window sizes (e.g., three-, four-, and five-channel windows) at different frequencies. Then, to search for candidates, we performed the source-detection algorithm (Section 2.3) on these moment-zero maps.

After having these candidates, we selected reliable targets. As mentioned previously, all O I absorbers have column densities of $\log(N_{\mathrm{OI}}/\mathrm{cm}^{-2}) > 14$. Guided by theoretical works, at $z \approx 6$, strong absorbers are generally linked to massive dark-matter (DM) haloes ($\log(M_h/M_\odot) \gtrsim 11$; Finlator et al. 2013; Keating et al. 2016), corresponding to stellar masses of $\log(M_*/M_\odot) \gtrsim 9$ (Ma et al. 2018). To search for such galaxies, we aim to detect [C II] emission from data cubes by constraining three key parameters: the line width of [C II] emission, the velocity offset, and the projected impact parameters between [C II] emitters and O I absorbers.

To begin with, following Le Fèvre et al. (2020), the line width of a reliable [C II] emitter at $z \sim 6$ should be $\gtrsim 250\,\mathrm{km\,s^{-1}}$, as measured by Capak et al. (2015). Similar line width justification is also reported in (e.g., Aravena et al. 2016; Fujimoto et al. 2019; Béthermin et al. 2020). Furthermore, the relative velocity offset between the absorber and its host galaxy should be within the range of $\pm 200\,\mathrm{km\,s^{-1}}$, because typical star-forming galaxies at $z \approx 6$ have been shown to have outflow velocities of $v \lesssim 200\,\mathrm{km\,s^{-1}}$ (e.g., Steidel et al. 2010; Keating et al. 2016; Díaz et al. 2021). We also note that due to projection effects, the projected velocity along the line of sight should be one component of the outflow velocity. As such, it is safe to constrain the velocity offset, our second constrained parameter, from $-200$ to $200\,\mathrm{km\,s^{-1}}$. For the projected impact parameter, cosmological simulations suggest that star formations and galactic outflows from star-forming galaxies are the primary mechanisms (Oppenheimer et al. 2009). These metal-enriched gases (particularly for strong absorbers, e.g., $\log(N_{\mathrm{OI}}/\mathrm{cm}^{-2}) > 14$) are reasonably gravitationally bounded in the circumgalactic medium (CGM) scale (Finlator et al. 2013; Keating et al. 2014). Guided by these theoretical models, the hosts of strong O I absorbers tend to reside in dark-matter halos with masses of $M_h \approx 10^{11-12}\,M_\odot$, corresponding to virial radii of <50 proper kpc (pkpc) at $z \sim 6$ (Keating et al. 2016). Therefore, the impact parameter between absorbers and hosts should be smaller than 50 pkpc. We select candidates with a fidelity level of 30%, which corresponds to the S/N $\geqslant 4$. To summarize, we constrain the following parameters

1. Line Width $\gtrsim 250\,\mathrm{km\,s^{-1}}$
2. $-200\,\mathrm{km\,s^{-1}} \leqslant$ Velocity Offset $\leqslant 200\,\mathrm{km\,s^{-1}}$
3. Impact parameter $\leqslant 50\,\mathrm{kpc}$
4. S/N $\geqslant 4$

We also discuss [C II]-emitter candidates with large velocity offset and impact parameters in Section 4.2.2.

After completing all three steps, except the first detection (called [C II]2054 below) reported by Wu et al. (2021), there are no emitters passing our criteria. In the case of nondetections, we used five-channel width windows, corresponding to the typical line width closing to $\sim 300\,\mathrm{km\,s^{-1}}$ (Aravena et al. 2016), centered on the expected frequency of [C II] emission to obtain final velocity-integrated flux maps. The detection and nondetection [C II] maps of our samples are shown in Figure 1. Although only one out of six showed a positive detection, we constrained the velocity-integrated flux ($S\Delta v_{\mathrm{[C\,II]}}$) of [C II] emitters to be within $3\sigma$ upper limits. Additionally, the [C II] luminosity ($L_{\mathrm{[C\,II]}}$) and the derived star formation rate (SFR$_{\mathrm{[C\,II]}}$; e.g., Wang et al. 2013; Schaerer et al. 2020) can be further constrained. The results are reported in Table 2. We conclude that, for the nondetection fields, the average $3\sigma$ upper limits of the various parameters are $S\Delta v_{\mathrm{[C\,II]}} < 0.06\,\mathrm{Jy\,km\,s^{-1}}$, $L_{\mathrm{[C\,II]}} < 5.8 \times 10^7\,L_\odot$, and SFR$_{\mathrm{[C\,II]}} < 5.5\,M_\odot\,\mathrm{yr^{-1}}$. Note that several emitters with larger impact parameters in Figure 1 are further discussed in Sections 4.2.2 and 4.3.

#### 3.1.2. Statistical Results

To constrain the galaxy absorber cross-correlation function, we constructed it as follows:

$$\xi_{\mathrm{g-abs}} = \frac{1}{n_0}\frac{\Delta N(r)}{\Delta V} - 1 = \left(\frac{r}{r_0}\right)^{-\gamma}. \quad (1)$$

In Equation (1), $n_0$ is the mean number density of galaxies, which can be obtained from the Luminosity function (LF) proposed by Loiacono et al. (2021). By integrating the LF to the bright end ($L > L_{\mathrm{[C\,II]\_obs}}$), we then obtain $n_0 = 1.3 \times 10^{-4}\,\mathrm{cMpc^{-3}}$. Similarly, $\Delta N(r)$ represents the number of galaxies in a spherical shell of survey volume $\Delta V$, where $r$ is the distance between an absorber and a galaxy. $r_0$ and $\gamma$ are the correlation length and power-law slope, respectively. Here, we recall the initial detection results. We have one [C II] emitter with SFR$_{\mathrm{[CIII]}} \approx 7\,M_\odot\,\mathrm{yr^{-1}}$ that is located approximately 20 kpc from OI-enriched gas at $z = 5.978$. Combined with our larger sample, however, only one [C II] emitter in six O I absorber fields is detected. Plugging in the final survey volume and one detected [C II] emitter into Equation (1) ($r = 20\,\mathrm{kpc}$ and $\Delta N(r) = 1$, and $\Delta V$ is the total survey volume of six fields), we get the relation between $r_0$ and $\gamma$. We add detailed calculations in Appendix B. This result is shown in the left panel of Figure 2. Although we only detect one [C II] emitter, the

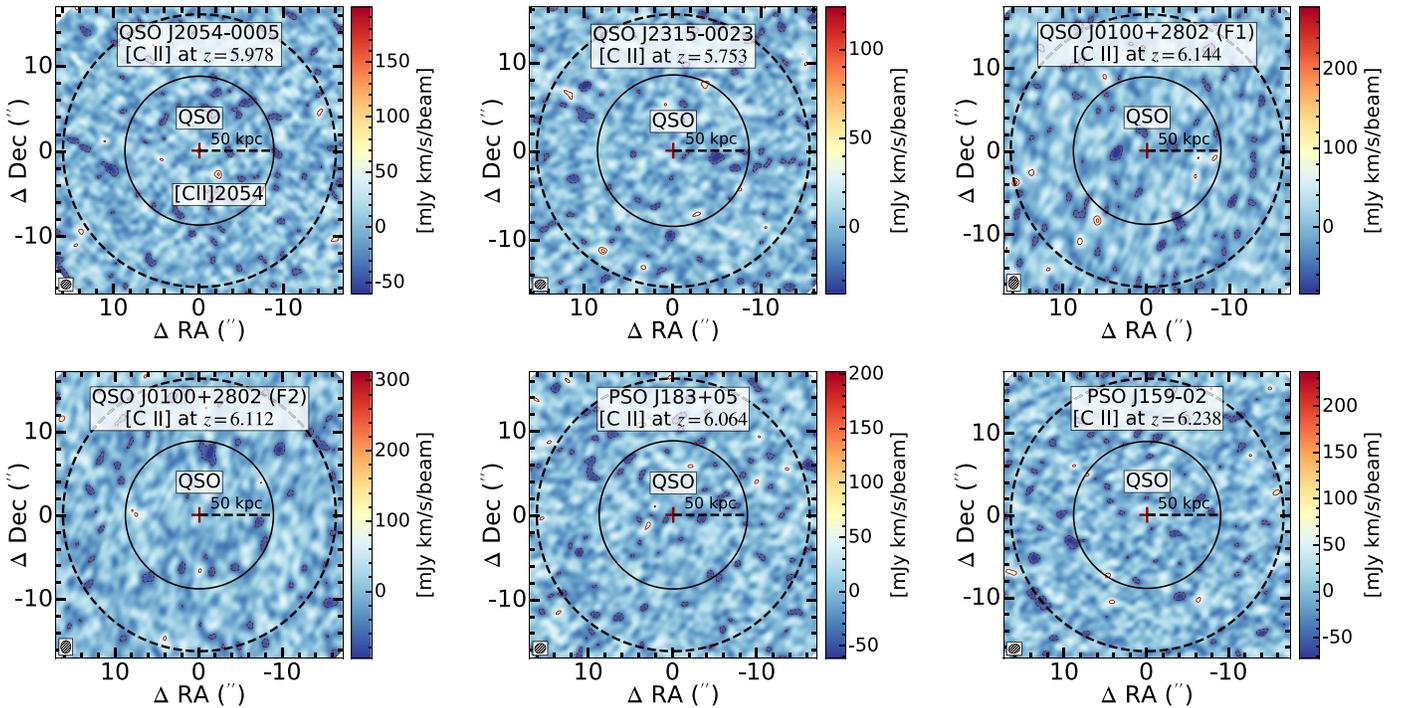

**Figure 1.** [C II] observations at the redshifts of six O I absorbers with $z = 5.978, 5.753, 6.144, 6.112, 6.064$, and $6.238$, respectively. Color bars represent the integrated [C II] flux. These moment-zero maps are integrated over the central $\sim 300$ km s$^{-1}$ at the O I absorber redshifts. Solid black lines are searching areas of [C II] candidates, while black dashed lines represent the regions with primary beam limits >20%. Outer red contours start at $3\sigma$, with increasing by $1\sigma$. Meanwhile, negative contours are dashed red lines and start at $-2\sigma$. The beam size of each moment-zero map is demonstrated in the bottom left corner.

**Table 2**
[C II] Moment-zero Map Properties

| Field Name | $S\Delta v_{\rm [C\,II]}$ (Jy km s$^{-1}$) | $L_{\rm [C\,II]}$ ($10^7\,L_\odot$) | SFR$_{\rm [C\,II]}$ ($M_\odot$ yr$^{-1}$) |
|---|---|---|---|
| (1) | (2) | (3) | (4) |
| J2054−0005* | $0.0758 \pm 0.0177$ | $7.0 \pm 1.7$ | $6.8 \pm 1.7$ |
| J2315−0023 | <0.03 | <2.7 | <2.5 |
| J0100+2802 (F1) | <0.07 | <7.1 | <6.8 |
| J0100+2802 (F2) | <0.08 | <7.9 | <7.6 |
| PSO J183+05 | <0.05 | <5.0 | <4.8 |
| PSO J159-02 | <0.06 | <6.2 | <5.9 |

**Notes.** Columns: (1) QSO field name; (2) integrated [C II] flux ($3\sigma$ upper limit); (3) [C II] luminosity; (4) [C II]-based star formation rate; ∗ means the field with detection and shows the results of the targets ([C II]2054).

observed $r_0$ and $\gamma$ relation is statistically greater than that of cosmological simulations (Finlator et al. 2020).

In addition to comparisons between the measured SFR and impact parameter, the properties of dark-matter halos that host O I absorbers can also strongly constrain cosmological simulations. Here, we do not directly compare the observed host halo properties with simulations because of the mismatching problem. It is inevitable to mismatch absorbers and their host galaxies in cosmological simulations. Usually, the host-halo-match algorithm is based on the closest match method, i.e., the metal-enriched gas is naturally assumed to be hosted by the nearest halo. However, massive halos can reasonably blow out metal-enriched gas far away, which may result in a relatively small halo being closer to the blownout gas. An example of this scenario can be found in Figure 16 in Keating et al. (2016). Therefore, we compare the number of galaxies clustered with O I absorbers to that predicted by simulations. In the right panel of Figure 2, we show how many galaxies are around O I absorbers within our given impact parameters. The gray line shows the results in Technicolor Dawn simulations (Finlator et al. 2020), while the blue line is a best-fit linear relation. Simulations suggest that the number of associated [C II] emitters is $\approx 10^{-2.5}$ within 50 kpc around O I absorbers. However, the identification of [C II]2054 suggests 1/6 of our O I absorber sample (red square), which is 1–2 orders of magnitude more abundant than that predicted by simulations. Considering the Poisson uncertainty at the 99% single-sided confidence level (Gehrels 1986), the observed number of [C II] emitters at our detection is $10^{-2.77}$ (red triangle). Given these results, although there is only one detected [C II] emitter in six absorber fields, the detection of such a bright [C II] emitter is unexpected. But, considering the lower limit, our results are also consistent with those predicted by simulations. More discussions on the detection and nondetection of [C II] emitters can be found in Section 4.2.

### 3.2. Continuum Observations

Our deep ALMA observations also yield several continuum-source detections. Continuum images of five QSO fields are shown in Figure 3. In this section, we further analyze the properties of these targets.

To obtain the number of continuum sources, we run the source-detection algorithm on the continuum images. The effective search areas of our data are defined by regions where the primary beam limit was higher than 20% and are represented by dashed lines in Figure 3. Again, sources with peak S/N higher than $4\sigma$ are regarded as detections. Here, we give a quick summary. Our five-field observations yield continuum images at 260.0, 286.6, 259.9, 262.4, and 256.0 GHz, respectively. The standard deviations of these

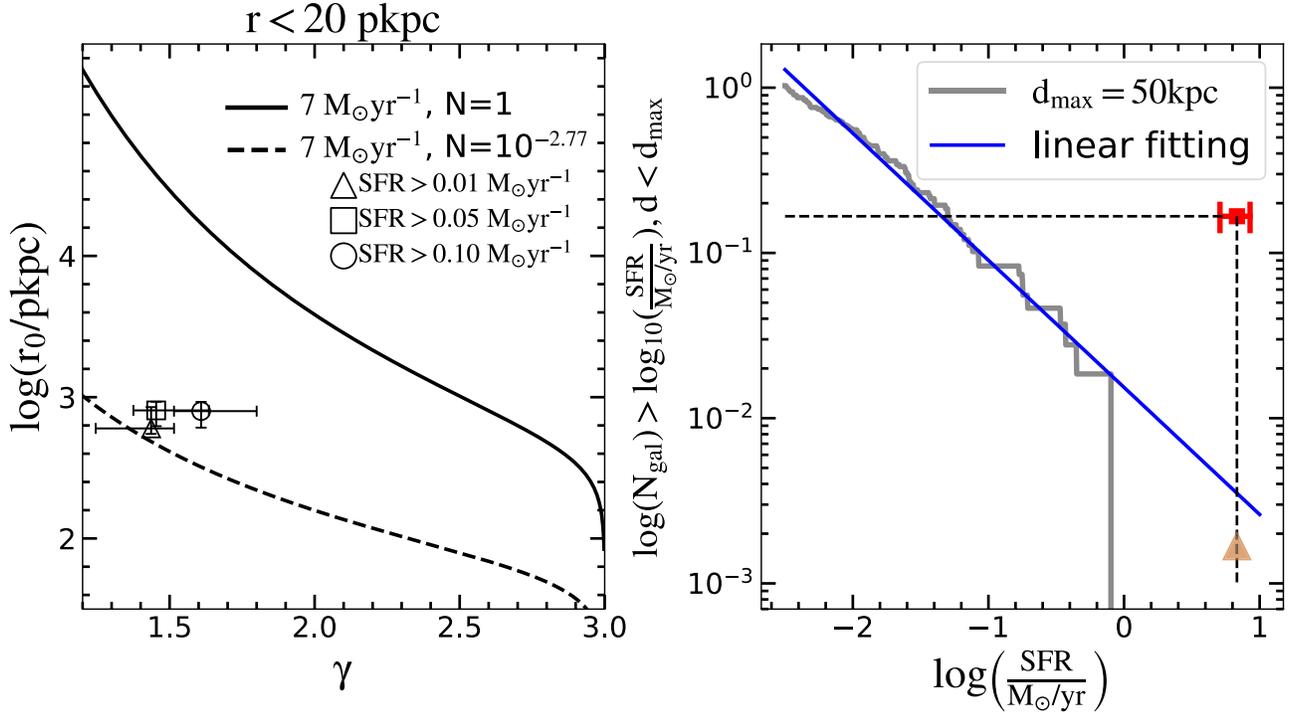

**Figure 2.** Left: Cross-correlation function between O I absorbers and their host galaxies. Symbols represent $r_0$ and $\gamma$ measured from the Technicolor Dawn simulation (Finlator et al. 2020) with $\log(N_{OI})$ comparable to our ALMA sample. Error bars are based on assuming the number of simulated host galaxies having Poisson fluctuations. The solid line shows the relation between $r_0$ and $\gamma$ based on one detected galaxy with SFR $\approx 7$ $M_\odot$ yr$^{-1}$ and within 20.0 pkpc in one of six absorber fields (see Section 3.1.2 for more details). The dashed black line shows the results derived from a 99% confidence-level Poisson uncertainty (Gehrels 1986) of the observed galaxy. Right: The cumulative number of galaxies around strong O I absorbers at $z \approx 6$. The solid line shows the number of galaxies as the function of SFR within impact parameters of 50 kpc from simulations (Finlator et al. 2020), while the blue line represents a best-fit linear relation. The red square represents the averaged number of galaxies in our six absorber fields, while the red triangle shows the lower limit according to the Poisson uncertainty at the 99% single-sided confidence level.

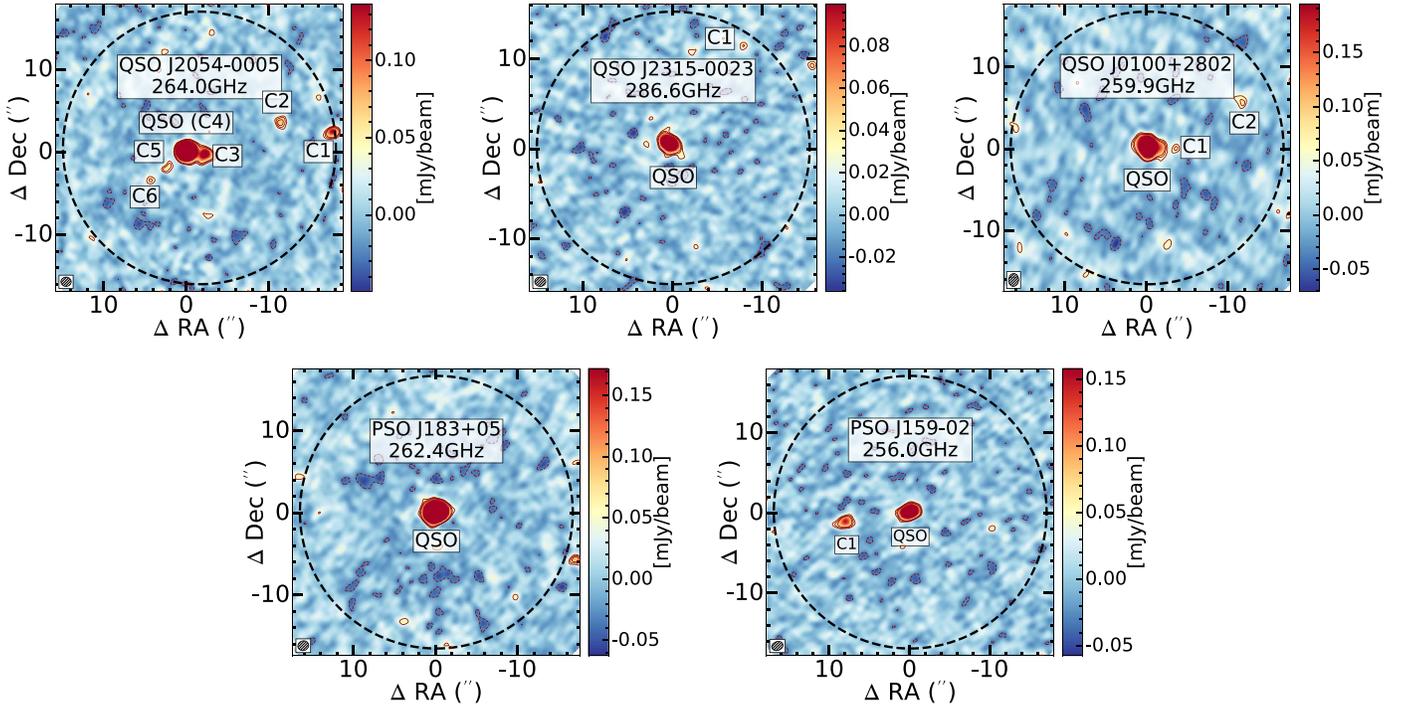

**Figure 3.** Continuum observations of five QSO fields at different frequencies. The $1\sigma$ rms of these fields are 0.012, 0.009, 0.018, 0.016, and 0.014 mJy/beam, respectively (from left to right). Contours are drawn at [3, 4, 5] $\times \sigma$. Dashed contours represent $-2\sigma$. Black dashed lines are marked as the boundary of the primary beam limit >20%. Beam sizes are plotted on the bottom left. Excluding five QSOs, we still detect nine continuum sources in these QSO fields. Detailed information is demonstrated in Table 3. Zoom-in figures are shown at the corner of each panel.

Table 3
Continuum Sources

| Target Name | R.A. (J2000) | Decl. (J2000) | $S_{\rm cont}$ (mJy) |
|---|---|---|---|
| (1) | (2) | (3) | (4) |
| J2054−0005C1 | 20:54:05.325 | −00:05:12.127 | 0.64 ± 0.06 |
| J2054−0005C2 | 20:54:05.742 | −00:05:10.890 | 0.12 ± 0.02 |
| J2054−0005C3 | 20:54:06.358 | −00:05:14.828 | 0.19 ± 0.01 |
| J2054−0005Q (C4) | 20:54:06.501 | −00:05:14.435 | 3.37 ± 0.01 |
| J2054−0005C5 | 20:54:06.649 | −00:05:16.482 | 0.067 ± 0.014 |
| J2054−0005C6 | 20:54:06.792 | −00:05:17.915 | 0.069 ± 0.016 |
| J2315−0023Q | 23:15:46.601 | −00:23:57.652 | 0.35 ± 0.01 |
| J2315−0023C1 | 23:15:46.055 | −00:23:46.697 | 0.13 ± 0.03 |
| J0100+2802Q | 01:00:13.021 | +28:02:25.822 | 1.27 ± 0.02 |
| J0100+2802C1 | 01:00:12.763 | +28:02:25.678 | 0.076 ± 0.019 |
| J0100+2802C2 | 01:00:12.156 | +28:02:31.593 | 0.18 ± 0.04 |
| PSO J183+05Q | 12:12:26.976 | +05:05:33.576 | 4.77 ± 0.02 |
| PSO J159−02Q | 10:36:54.184 | −02:32:37.938 | 0.63 ± 0.01 |
| PSO J159−02C1 | 10:36:54.718 | −02:32:39.164 | 0.21 ± 0.02 |

**Notes.** Columns: (1) name of continuum source, where QSO host galaxies are appended by a Q after the field name, while continuum sources are appended by a C; (4) continuum flux density.

fields are 0.012, 0.009, 0.018, 0.016, and 0.014 mJy/beam. Excluding the five identified QSOs, there are nine sources detected with a mean flux density of 0.19 mJy and S/N of 7.6. The sources catalog is presented in Table 3. We note that these newly detected continuum sources in our observations have no redshift information.

## 4. Discussion

### 4.1. Continuum-source Number Counts

To evaluate the number-count excess of continuum sources in our observed fields, we compare observed continuum-source numbers with those in random fields.

Our continuum observations are conducted around ∼266 GHz, corresponding to ∼1.1 mm. To date, there are a handful of deep submillimeter-continuum surveys that conduct random-field 1.1 mm continuum galaxy number counts (e.g., Fujimoto et al. 2016; Franco et al. 2018; González-López et al. 2020). To reveal the continuum-source excesses in our QSO fields, we compared the number of detected sources to the expected number in blank fields based on the 1.1 mm galaxy number counts. To do a proper analysis, our observed flux densities at different frequencies need to be converted to flux densities at 1.1 mm. We do this conversion by assuming a modified blackbody emission model, specifically, $S_\nu \propto \nu^{(3+\beta)}/(\exp(h\nu/kT_{\rm dust}) - 1)$ (Popping et al. 2020). For this model, we adopt $\beta = 2.0$ and $T_{\rm dust} = 38$ K for star-forming galaxies at $z \sim 6$ (Faisst et al. 2020). For QSO field, J2054, J2315, J0100, PSO J183 and PSO J159, the flux converted ratios are $S_{1.1\,{\rm mm}}/S_{264.0\,{\rm GHz}} = 1.13$, $S_{1.1\,{\rm mm}}/S_{286.6\,{\rm GHz}} = 0.83$, $S_{1.1\,{\rm mm}}/S_{259.9\,{\rm GHz}} = 1.20$, $S_{1.1\,{\rm mm}}/S_{262.4\,{\rm GHz}} = 1.16$, and $S_{1.1\,{\rm mm}}/S_{256.0\,{\rm GHz}} = 1.27$, respectively. For these models, the radii for the efficient searching area in these five fields are 16″.5, 15″.2, 16″.7, 16″.6, and 17″.0. We note that the efficient searching areas are defined as primary beam limits >20%. From this analysis, we obtain the completeness-corrected continuum-source densities in five QSO fields (see Figure 4). We list details of completeness correction in Appendix B.

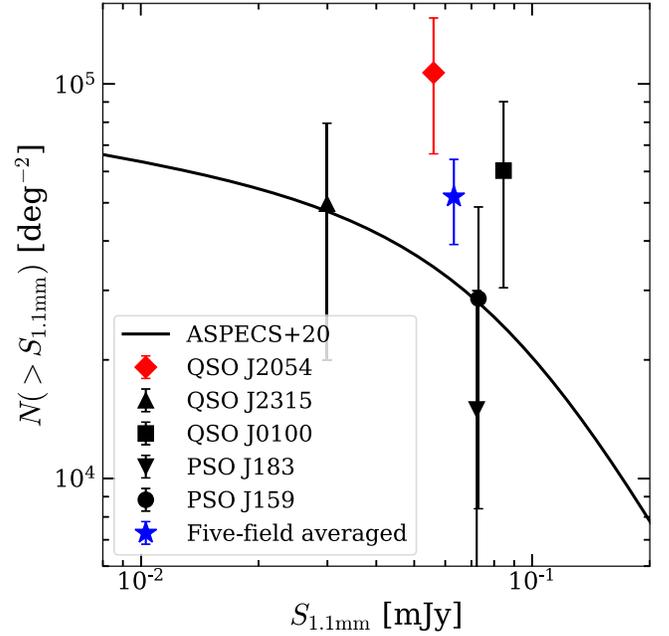

**Figure 4.** The number function of 1.1 mm continuum sources. The black symbols represent the number of sources in our QSO fields. The red diamond shows the results in the field of QSO J2054-0005. Error bars are determined by assuming Poisson uncertainties. The blue star is the five-field-averaged result. The black line is the double power-law fitted results from González-López et al. (2020).

In Figure 4, the different black symbols represent the observed continuum-source densities in different fields, while the blue star represents the five-field-averaged result. The measurement in the field of QSO J2054-0005 is shown in a red diamond. Error bars are estimated based on the Poissonian noise of the observed number of continuum sources. Our results suggest that within the field of view of ALMA, only the QSO J2054−0005 field shows a higher number density of continuum sources than the prediction from the continuum number counts function (González-López et al. 2020; Popping et al. 2020). Our observations are also consistent with that described in Champagne et al. (2018). However, we note that, due to the size of the ALMA field of view, one-pointing observations may miss the true number of excesses on the scale of ≳30″. Meyer et al. (2022) conducted multidithering ALMA observations to map the environments surrounding QSOs and also found an overabundance case of continuum sources in one of three QSO fields. This scenario is explained by a possible foreground overdensity caused by galaxies at the cosmic noon. This effect could also be suitable to the excess of dusty sources in J2054-0005. Although we only identify one overabundance, we conclude that more large-scale continuum observations in QSO fields will be necessary for future analysis of the environment around QSOs at high redshift or galaxy overdensities at any other redshift.

### 4.2. Detection and Nondetection of [C II] Emitters

#### 4.2.1. Gravitational Lensing Effects

It is intriguing that we only have one detection of a [C II] emitter. In this section, we use our previous discussion and our overdensity measurements to propose one possible explanation.

In this field, we find that the observed continuum sources are ∼3× as abundant as the random field (Figure 4). As noted in Wu et al. (2021), these continuum sources are foreground galaxies at $z \sim 2$–4. Thus, in this section, we build up lensing models and discuss the details of the possibilities of lensing-caused detection.

To reveal the possibility of galaxy-lensing-caused detection, we demonstrate one rough approach by building up a lensing model. There are three parameters of foreground galaxies: redshifts, halo masses, and concentrations. We first use five HST broadband observations (Wu et al. 2021) to estimate the photometric redshift of the foreground targets. We ran `eazy-py`[14] with flat priors and found most of them located at the redshift of $z \sim 1.75$. Then, the halo masses are converted based on the stellar mass–halo mass relation (Behroozi et al. 2010). We estimate the stellar masses of these targets using `CIGALE` (Boquien et al. 2019). The averaged stellar mass is $M_* \approx 10^{9.6} M_\odot$, corresponding to a halo mass of $M_h \approx 10^{11.4} M_\odot$. Next, the concentrations of host halos of these galaxies are estimated based on the mass-concentration relation (Klypin et al. 2016; Ludlow et al. 2016; Child et al. 2018). The mean value of concentration is $c \approx 5.9$. Finally, the lensing model is calculated under the assumption of NFW profiles using `lenstronomy`[15] (Birrer & Amara 2018; Birrer et al. 2021). After constructing a lensing model, we found that, at the location and redshift of our first detection, the magnification $\mu \approx 1.1$. We conclude that, under the rough calculations shown above, the first detection is not caused by the galaxy-caused gravitational lensing effects.

*4.2.2. [C II] Emitter Candidates with Larger Impact Parameters and Higher Velocities*

Except for the field J2054, our other field observations yield nondetection results. Thus, we discuss the scenarios that caused by more [C II] candidates satisfying the following conditions. Díaz et al. (2021) reported LAEs as C IV absorber host galaxies having a velocity offset of ∼600 km s$^{-1}$ and impact parameters larger than 50 kpc (Figure 15 and Table 2 in their manuscript).

1. Line Width $\gtrsim 250$ km s$^{-1}$
2. $-600$ km s$^{-1} \leqslant$ Redshift Offset $\leqslant 600$ km s$^{-1}$
3. Impact parameter $\leqslant 100$ kpc
4. S/N $\geqslant 4$

Adopting these conditions, in six O I absorber fields, we find 14 candidates with $4 <$ S/N $< 5$, and no sources having S/N $\geqq 5$. The averaged S/N of these candidates is ≈4.3. The [C II] luminosities of these targets are in the range of 5.4–47.3 $L_\odot$, while the range of impact parameters is ∼24.2–92.1 kpc. To further analyze the reliability of these candidates, we measured the fidelity of our data by comparing the number of positive and negative targets. More details are shown in Appendix A. At the S/N of ≈4.3 level, the fidelity is close to 30%, which means that there are four real targets within our 14 candidates. We put detailed information on these candidates in Appendix C. We note that future observations (e.g., JWST) could help to validate the reality of these candidates by checking their rest-frame optical emission lines (Bordoloi et al. 2023; Kashino et al. 2023).

---

[14] https://github.com/gbrammer/eazy-py
[15] https://github.com/sibirrer/lenstronomy

*4.3. The Role of Outflow Velocities*

We analyze the impact parameters again to discuss the physics behind our observations. Theoretical work predicts that metal-line absorbers are blown out by galactic feedback (Muratov et al. 2015; Doughty & Finlator 2019; Finlator et al. 2020). In an effort to test the efficiency of galactic winds in transporting metals in our sample, following Díaz et al. (2021) and Galbiati et al. (2023), we assume the outflow starts at redshifts higher than those of the observed metal absorbers ($z = 7$–10, Galbiati et al. 2023). We then compare the observed and the expected impact parameters between galaxies and absorbers. The predicted impact parameters are calculated based on the assumption of the projected velocity of the galactic wind ($\langle v(z) \rangle$) as the following equation:

$$\text{impact parameter} = \langle v(z) \rangle \times [t_{z\_\text{start}} - t(z)], \quad (2)$$

where $t_{z\_\text{start}}$ is the lookback time at the launching redshift of galactic outflow, and $t$ is the lookback time at the observed redshifts.

This comparison allows us to test different feedback models (Keating et al. 2016) with different outflow velocities. Díaz et al. (2021) regarded LAEs that are located near C IV absorbers as the real absorber host galaxies and found eight C IV-LAE pairs at $z = 5$–6 in MUSE observations. Their results suggest that the $\langle v \rangle$ values of metal-line absorber-associated galaxies mostly range from 100–150 km s$^{-1}$. In Figure 5, we demonstrate the expected impact parameters as the combinations of different $t_{z\_\text{start}}$, and $\langle v \rangle$. The gray-colored star and unfilled stars are our samples. We note that outflows launched more recently before the observations will be linked to more intense star formation activity, providing strong galactic winds with large velocities.

## 5. Summary

We present deep ALMA [C II] and continuum observations in five QSO fields containing six strong O I absorbers at $z \sim 6$. By performing a source-finder algorithm on the [C II] moment-zero and continuum maps, we obtain nondetections of [C II] emitters, while nine additional, new continuum sources are detected. Our findings are summarized as follows:

1. If we constrain the velocity range of $\pm 200$ km s$^{-1}$ and impact parameter of 50 kpc, our ALMA observations yielded nondetections of five strong O I absorption ($\log(N_{\text{OI}}/\text{cm}^{-2}) > 14$) associated galaxies. Including the detection we reported in Wu et al. (2021), we have one detection per six O I absorber fields. Further, we also found 14 tentative detections with S/N of ≈4.3 and an averaged SFR of 17.4 $M_\odot$ yr$^{-1}$ at large impact parameters (>50 kpc) and having larger velocity offsets of within $\pm 600$ km s$^{-1}$. If these detections are confirmed in the future, then the scenario that massive galaxies blow out metals in larger distances with higher velocities may be favored from the theoretical side.

2. Although we only detected one [C II] emitter, the detection rate of bright sources is still much higher than that predicted from simulations. The observed correlation length ($r_0$) and power-law index $\gamma$ of the galaxy absorber correlation function $\xi_{\text{g-abs}}$ is 1–2 orders higher than that measured from cosmological simulations Finlator et al. (2020). Meanwhile, comparing the cumulative number of galaxies around absorbers with that in cosmological

 

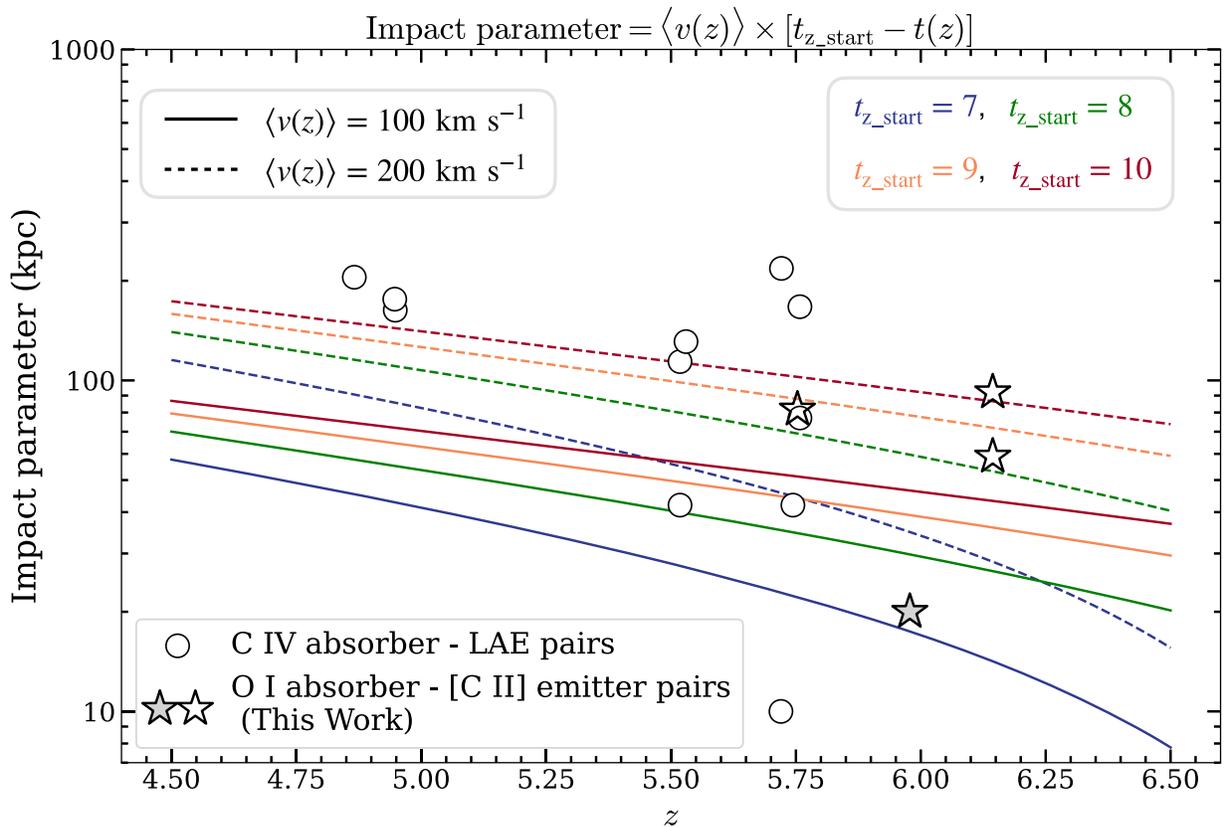

**Figure 5.** The impact parameter between absorber-galaxy pairs. The unfilled dots are C IV absorber–LAE pairs reported by Cai et al. (2017) and Díaz et al. (2021). The filled gray star is the identified O I absorber-[C II] emitter pair, while the nonfilled stars are emitters in our data cube but get ruled out because of large impact parameters. The solid and dashed lines are calculated based on the projected outflow velocities of 100 and 200 km s$^{-1}$, respectively. Different colors indicate different launching times of galactic outflow ranging from $z = 7$–10.

simulations, the detected [C II] emitter is brighter at a factor of 10 than the expected one in simulations.

3. Using the method proposed by Díaz et al. (2021) and Galbiati et al. (2023), we use the measured impact parameter to test the mean speeds $\langle v \rangle$ of galactic winds, based on different assumptions of the launching time of galactic outflows. For the typical outflow velocities of galactic wind ($\sim$100–200 km s$^{-1}$), at the observed redshift, a more recently launched outflow will be more intense to eject the metal-enriched gas out to the observed impact parameter.

4. ALMA observations provide us with deep continuum observations with a field of view with a diameter of $\sim$30″. No significant continuum-source number excesses are observed in QSO fields other than J2054. Our results suggest that QSOs do not directly trace overabundances, which is also consistent with Champagne et al. (2018). Nevertheless, future observations are required to further confirm the overabundances on a large scale (Meyer et al. 2022).

Absorbers discovered from QSO sightlines can directly trace gas that provides the fuel for star formation and regulates the formation of galaxies. Yet ALMA observations have provided constraints to early CGM/IGM enrichment theoretical models up to $z \approx 6$. Our pilot results also suggest that future James Webb Space Telescope (JWST) observations of absorber-galaxy interaction at the epoch of reionization are necessary. We aim to use JWST, as JWST also plans to observe the same sightline with the slitless spectroscopic mode, which will enable us to detect the [O III] emission $\lambda\lambda$4959, 5007 associated with the [C II] emitter.

### Acknowledgments

Y.W. thanks Wenshuo Xu, Xiaojing Lin, and Mingyu Li for the fruitful discussion. Z.C., Y.W., and S.Z. are supported by the National Key R&D Program of China (grant No. 2018YFA0404503) and the National Science Foundation of China (grant No. 12073014). The science research grants from the China Manned Space Project with No. CMS-CSST2021-A05, and Tsinghua University Initiative Scientific Research Program (No. 20223080023). The Cosmic Dawn Center is funded by the Danish National Research Foundation. K.F. gratefully acknowledges support from STScI Program #HST-AR-16125.001-A. This program was provided by NASA through a grant from the Space Telescope Science Institute, which is operated by the Associations of Universities for Research in Astronomy, Inc., under NASA contract NAS5-26555. K.F.'s simulation utilized resources from the New Mexico State University High Performance Computing Group, which is directly supported by the National Science Foundation (OAC-2019000), the Student Technology Advisory Committee, and New Mexico State University and benefits from inclusion in various grants (DoD ARO-W911NF1810454; NSF EPSCoR OIA-1757207; Partnership for the Advancement of Cancer Research, supported in part by NCI grants U54 CA132383 (NMSU)). M.N. acknowledges support from ERC

Advanced grant 740246 (Cosmic_Gas). F.W. is thankful for the support provided by NASA through a NASA Hubble Fellowship (grant No. HST-HF2-51448.001-A) awarded by the Space Telescope Science Institute, which is operated by the Association of Universities for Research in Astronomy, Inc., under NASA contract NAS5-26555. The National Radio Astronomy Observatory is a facility of the National Science Foundation operated under cooperative agreement by Associated Universities, Inc. L.K. was supported by the European Unions Horizon 2020 research and innovation program under the Marie Skodowska-Curie grant agreement No. 885990. For the purpose of open access, the author has applied a Creative Commons Attribution (CC BY) license to any Author Accepted Manuscript version arising from this submission.

*Facility:* ALMA.

*Software:* astropy (Astropy Collaboration et al. 2022), lenstronomy (Birrer & Amara 2018; Birrer et al. 2021), CASA v.5.6.1-8 (McMullin et al. 2007), DAOStarFinder (Bradley et al. 2020), CIGALE (Boquien et al. 2019), Qubefit (Neeleman et al. 2020), EAZY-py (Brammer et al. 2008), interferopy (Boogaard et al. 2021).

## Appendix A
## Fidelity

To get the fidelity of our [C II] emitter detection, we use *interferopy* to search for both positive and negative targets in six O I absorber fields. The blue and orange lines in the left panel of Figure 6 show the cumulative distribution of positive and negative targets. To avoid the small number statistics at the tail of the distribution, following González-López et al. (2019), we regarded the number of negative sources as a function with the form of $N = 1 - \mathrm{erf}(\frac{S/N}{\sqrt{2}\sigma})$, where erf is the error function. We then integrated this function into the low-S/N end to get the cumulative distribution (red line in the left panel). Following Aravena et al. (2016), we define the fidelity $P$ as $P(>S/N) = 1 - \frac{N_{\mathrm{negative}}}{N_{\mathrm{positive}}}$, where $N_{\mathrm{positive}}$ and $N_{\mathrm{negative}}$ are the cumulative number of positive and negative targets, respectively. The measured values are shown in gray bars in the right panel of Response Figure 6. We also fitted the measured fidelity as a function of S/N with the form of $P = 1/2 + \mathrm{erf}(\frac{S/N - c}{\sigma}))/2$. The fitted result is shown as the dark-red line.

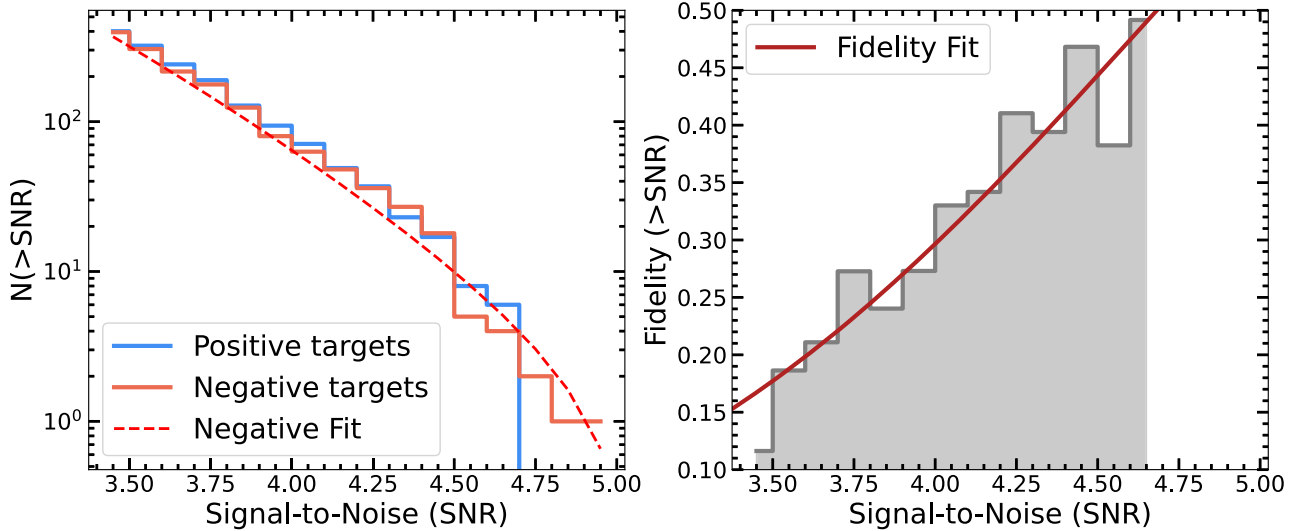

**Figure 6.** Left: Cumulative number distribution of emitters in six O I absorber fields as the function of S/N. Blue and orange lines are the measured numbers in six fields, respectively. The red dashed line shows a fitted error function to the number of negative targets. This fitting is in order to avoid the small number of statistics at the high-S/N end. Right: The measured fidelity as the function of S/N. The gray bars show the measured fidelity. The fitted fidelity is shown in dark red.

## Appendix B
## Completeness Correction for Continuum Sources

To estimate the actual number of sources in our observations, we insert pseudo-continuum targets into each image of each field and then obtain the detection completeness. Details of the completeness estimation mostly follow the procedure described in Fujimoto et al. (2016), Franco et al. (2018), and Gómez-Guijarro et al. (2022).

We first mask continuum sources with S/N > 4 to avoid the contamination caused by tentative detections. The pseudo-continuum targets were generated using the flux-scaled synthesized beam with S/N ranging from 4 to 6 in steps of 0.1. For each S/N bin, we randomly insert 100 sources into each image and re-perform the source-detection algorithm. If the detected location of an artificial source is close to that inserted within a distance smaller than one beam size, we regard this detection as recovered. The completeness of these QSO fields is shown in Figure 7. The red line is fitted by the function of $1 - \exp(aS/N - b)$. We find that sources with S/N ranging from 4–5 have a mean probability of being detected of ~70%. The completeness-corrected number of continuum sources in fields J2054, J2315, J0100, PSO J183, and PSO J159 are 7.0, 2.8, 4.1, 1.0, and 2.0, respectively.

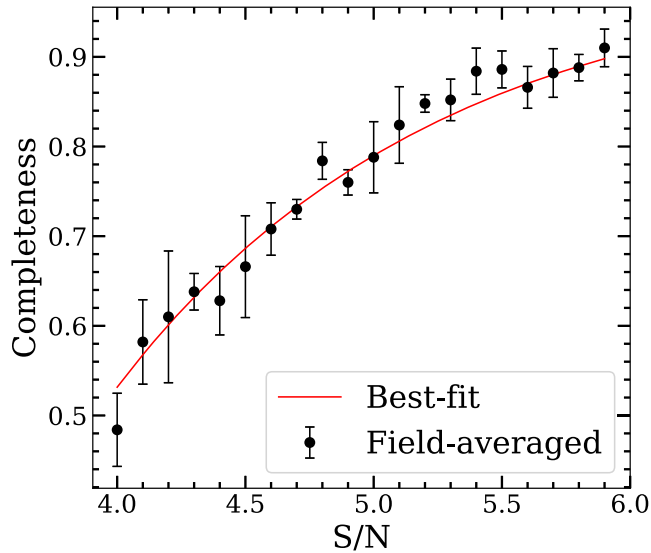

**Figure 7.** Completeness as a function of S/N. Circles are the field-averaged completeness, while error bars represent the field-to-field variance in five QSO fields. The red line is the best-fit function.

## Appendix C
## Correlation Length $r_0$ and Power-law Index $\gamma$ Calculation

In this section, we show the detailed calculations of the correlation length and power-law index. The correlation function is defined as the number excess of galaxies around one metal absorber in a given survey volume.

$$\xi_{\text{g-abs}} = \frac{1}{n_0}\frac{\Delta N(r)}{\Delta V} - 1. \text{ Thus we have,}$$

$$n_0[1 + \xi_{\text{g-abs}}(r)] = \frac{\Delta N(r)}{\Delta V}. \quad \text{(C1)}$$

If we then assume a power-law function, $\xi_{\text{g-abs}} = \left(\frac{r}{r_0}\right)^{-\gamma}$, we can then replace $\xi_{\text{g-abs}}$ as a power law and get the following equation:

$$n_0\left[1 + \left(\frac{r}{r_0}\right)^{-\gamma}\right] = \frac{\Delta N(r)}{\Delta V}. \quad \text{(C2)}$$

We assumed a 3D spherical survey volume with a given radius of $r$. We then integrate this equation and obtain:

$$4\pi \int_0^r n_0\left[1 + \left(\frac{r}{r_0}\right)^{-\gamma}\right] = N(r). \quad \text{(C3)}$$

To justify the spherical approach, we use the following assumptions proposed by simulations. In simulations (Keating et al. 2016), strong O I absorbers are always included within the DM halo of galaxies. For the successfully detected system, the halo mass is $4 \times 10^{11} M_\odot$ (Wu et al. 2021), corresponding to the virial radius of ~20–30 pkpc at $z \approx 6$. Further, the observed impact parameter (~20 pkpc) is very close to the halo virial radius. Thus, to include this system in the DM halo, the line-of-sight distance will be small. A sphere-like survey volume with a radius of 20 pkpc will thus be large enough to include this target. In this work, we detect one galaxy at 20 pkpc. Thus, $N$ (20 pkpc) = 1. We then obtained:

$$4\pi \int_0^r n_0\left[1 + \left(\frac{r}{r_0}\right)^{-\gamma}\right] = 1, \text{ where } r = 20 \text{ pkpc.} \quad \text{(C4)}$$

By integrating this equation, we then obtain a relation between $r_0$ and $\gamma$:

$$\log(r_0) = \frac{\log[A(\gamma)]}{\gamma}, \text{ where}$$

$$A \equiv \left(\frac{1}{4\pi n_0} - \frac{r^3}{3}\right) \times \left(\frac{3-\gamma}{r^3}\right), \text{ where } r = 20 \text{ pkpc.} \quad \text{(C5)}$$

## Appendix D
## Velocity Offsets versus Impact Parameters of [C II] Candidates

In this section, we show the comparison between our two sets of criteria. Our major criteria are guided by theoretical simulations and described in Section 3.1.1. Cosmological simulations suggest that these metal absorbers are mostly generated by star formations and galactic outflows (Oppenheimer et al. 2009). Therefore, the velocity offset between absorbers and the actual hosts should be close to the outflow velocity (~200 km s$^{-1}$, Steidel et al. 2010). Further, these strong absorbers need to be gravitationally bounded in the circumgalactic medium (CGM) scale (Keating et al. 2014). The impact parameters thus need to be smaller than the virial radii of DM halos (<50 pkpc, at $z \sim 6$ Keating et al. 2016). The criteria are shown in the dark-gray region in Figure 8. We note that galaxies in the dark-gray region could be more physically connected to the O I absorbers.

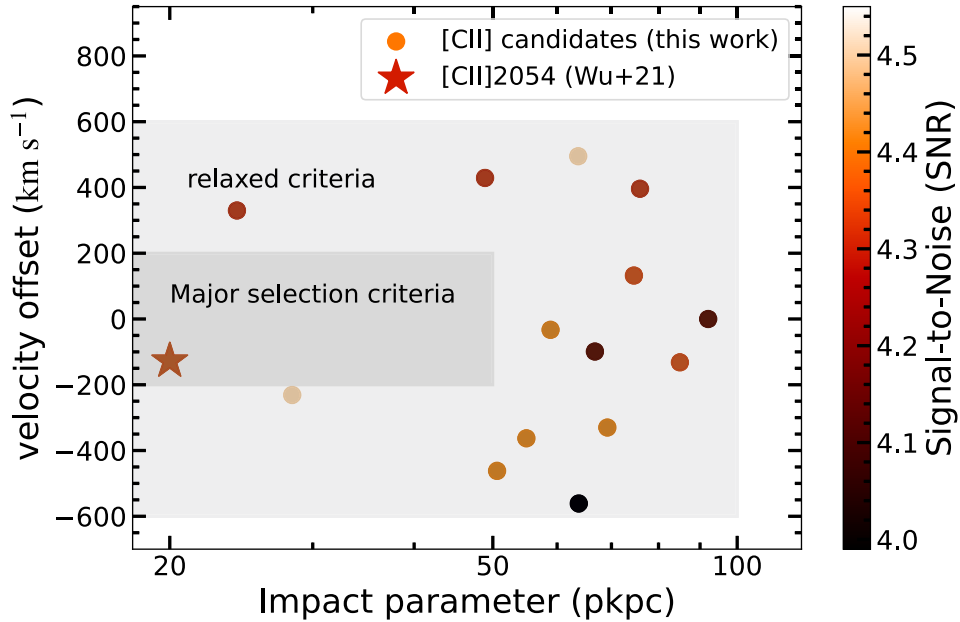

**Figure 8.** Velocity offsets vs. impact parameters of [C II] candidates in this work. These targets are all color-coded by their signal-to-noise ratios. The dark and light gray regions show our major and relaxed selection criteria, respectively.

## ORCID iDs

Yunjing Wu ● https://orcid.org/0000-0003-0111-8249
Zheng Cai ● https://orcid.org/0000-0001-8467-6478
Jianan Li ● https://orcid.org/0000-0002-1815-4839
Kristian Finlator ● https://orcid.org/0000-0002-0496-1656
Marcel Neeleman ● https://orcid.org/0000-0002-9838-8191
J. Xavier Prochaska ● https://orcid.org/0000-0002-7738-6875
Bjorn H. C. Emonts ● https://orcid.org/0000-0003-2983-815X
Shiwu Zhang ● https://orcid.org/0000-0002-0427-9577
Feige Wang ● https://orcid.org/0000-0002-7633-431X
Jinyi Yang ● https://orcid.org/0000-0001-5287-4242
Ran Wang ● https://orcid.org/0000-0003-4956-5742
Xiaohui Fan ● https://orcid.org/0000-0003-3310-0131
Emmet Golden-Marx ● https://orcid.org/0000-0001-5160-6713
Laura C. Keating ● https://orcid.org/0000-0001-5211-1958
Joseph F. Hennawi ● https://orcid.org/0000-0002-7054-4332